\documentstyle[prl,aps,psfig]{revtex}
\begin{document}
\draft

\twocolumn[\hsize\textwidth\columnwidth\hsize\csname
@twocolumnfalse\endcsname

\widetext
\title{ Charge fluctuations close to phase separation
 in the two dimensional $t-J$ model }
\author {  Matteo Calandra, Federico Becca  and Sandro Sorella }
\address{ 
 Istituto Nazionale di Fisica della Materia and
  International School for Advanced Studies 
Via Beirut 4, 34013 Trieste, Italy 
  } 
\date{\today}
\maketitle
\begin{abstract}
We have studied the t-J model   using the Green Function 
Monte Carlo  technique. 
 We have obtained   accurate energies well converged 
 in the thermodynamic  limit, by performing simulations up to
 $242$ lattice sites.   
  By studying the energy as a function of hole doping 
we conclude that   there is no phase separation in the physical region, 
relevant for HTc superconductors.  
This finding is further supported by the  hole-hole correlation function 
calculation.
  Remarkably, by approaching  the phase separation instability,
 for $J_c/t\sim 0.5$,
 this function displays enhanced fluctuations at incommensurate 
wavevectors, scaling linearly with the doping, in agreement with experimental 
findings. 
\end{abstract}
]
\narrowtext
 One of the most important questions raised  in the context of 
High-Tc superconductivity is whether the strong competition between 
hole propagation and antiferromagnetic (AF) order in the $CuO_2$  planes,
leads to   the segregation of  holes in regions without   AF order 
 (phase separation)\cite{jor}  or,  analogously,  to  
the formation of one dimensional ''stripes''  
with high hole-density\cite{tranquada}.

The simplest model taking into account spin interactions and hole kinematics 
is the $t-J$ model \cite{zhang}
\begin{equation}
\label{tjmodel}
H = J \sum_{\langle i, j \rangle} ( S_i \cdot S_j - \frac{1}{4} n_i n_j ) 
- t \sum_{\langle i, j\rangle , \sigma} ( c^{\dagger}_{i \sigma}  
c_{j \sigma} + h.c. ),
\end{equation}
where $c^{\dagger}_{i \sigma}$ creates an electron  of spin $\sigma$ on 
the site $i$, $n_i$ and $S_i$ being the electron number and spin operators
respectively. Double occupations  are  not allowed  
and summations are extended to nearest neighbors.

Following \cite{emery}, whenever  
 the ground state energy per hole $e_h(\delta)$
as a function of the doping $\delta$
\begin{equation}  \label{ex}
e_h(\delta)= (e(\delta)-e_0)/\delta 
\end{equation}
has a minimum for a non zero value of $\delta=x_c$ 
phase separation (PS) is energetically stabilized. 
In fact, by using the 
Maxwell construction,  a macroscopic gain in energy can be obtained by 
phase separating the holes at all doping $\delta < x_c$  in
 a hole rich phase with $\delta=x_c$ and a fully AF  region without holes 
with energy $e_0=-1.16944(4)J$\cite{calandra,sandvik}.

 Many authors\cite{emery,lee,manousakis,kohno,heeb}
  have tried to clarify
 the important issue of PS by using  numerical techniques.
Calculations of exact ground state 
energies on small lattice sizes\cite{emery}   do confirm the existence of PS 
at physical  $J/t$ and doping  
, but the sizes considered are far to be 
representative of the thermodynamic limit, where the question is meaningful.
By contrast,  using  the high temperature expansion  (HTE), 
 $PS$ was found only for  large $J/t$\cite{rice}.  

Much more information, not only on PS,  can be obtained  by studying the
 behavior of the hole - hole correlation function:
\begin{equation}
N(q)=\frac{1}{L} \sum_{i,j}\,e^{iq(R_i-R_j)}\,(1-n_{i})(1-\,n_{j})
\end{equation}
where $L$ is the number of lattice sites. The onset of $PS$ is characterized
 by the divergence of $N(q)$ for small momenta $q$. 
According to the HTE\cite{putikka}, $N(q)$ displays some interesting 
feature at twice the spinless fermion  
Fermi vector but no  PS in the physical region.  
These  results  are not affected 
by small size effects  but  are obviously limited by   the   difficulty  
of extrapolating    an  high temperature series  to zero temperature.

In this letter we 
tackle the PS problem 
by the lattice Green Function Monte Carlo (GFMC)
technique.
 This method  
allows to  filter out from a given trial wavefunction
$|\psi_T\rangle$ the ground state   $|\psi_0\rangle$
 of $H$ by statistical application of the power method:
\begin{equation}\label{iter}
|\psi_0\rangle  \propto (\Lambda -H)^n |\psi_T\rangle,
\end{equation}
where $\Lambda$ is a  suitably large constant required to allow 
convergence to the ground state for large $n$.
 In this scheme a Markov process is defined, acting on electron configurations 
 $\{x\}$  with definite positions and spins, which are changed  according 
to the Green Function 
$G_{x^\prime,x} = \Lambda \delta_{x^\prime,x} - H_{x^\prime,x}$\cite{sorella}.
The algorithm is efficient even for large system size, provided all 
the $G_{x^\prime,x}$  are non negative, otherwise one is facing the well 
known ''sign problem''.
 For the diagonal elements the above condition can be easily fulfilled 
by increasing $\Lambda$. 
However  an exceedingly large value of $\Lambda$ 
-which is often the case especially  for fermions- determines 
 a slowing down of the algorithm, since  
 there is a very small probability $\sim 1/\Lambda$ to accept 
a new configuration $x^\prime$ and   the algorithm 
remains almost always stacked in 
the old one $x$. Thus one needs much more power iterations (\ref{iter}) 
to generate statistically independent configurations. 
In order to overcome this difficulty, following \cite{trivedi}, it is better 
to determine {\em a priori} the number of diagonal moves before 
an off diagonal is accepted. Thus one can generate  each time a new
 configuration  without caring of a very large value of $\Lambda$.
A further improvement is to let $\Lambda \to \infty$ with $n =\Lambda \tau$, 
so that Eq.(\ref{iter}) becomes a more conventional imaginary time 
evolution by the operator $e^{ - H \tau}$. Further details of this technique
will be given elsewhere.

The negative off-diagonal elements of the Green Function can be handled 
approximately by a recent  achievement in GFMC:
 the fixed node technique (FN) for  lattice Hamiltonians\cite{ceperley}.
In order to control and improve the accuracy of the FN 
approximation in a systematic way, we have used a recent development, the 
GFMC with stochastic reconfiguration  (GFMCSR)\cite{sorella}. 
 The GFMCSR results presented here have been obtained by 
keeping  unchanged before and after each stochastic reconfiguration   
the energy, the kinetic energy and  the nearest neighbor hole-hole correlation
function.     GFMCSR   has to be implemented  with a very large number  $M$ 
of walkers   in   order to  stabilize the simulation ($M \sim 2000 \div 4000$), 
whereas in the FN calculations $M=800$ is by far  sufficient to control 
the bias due to the finite walker population with the technique described  
in Ref. \cite{calandra}.
\begin{figure}
\centerline{\psfig{figure=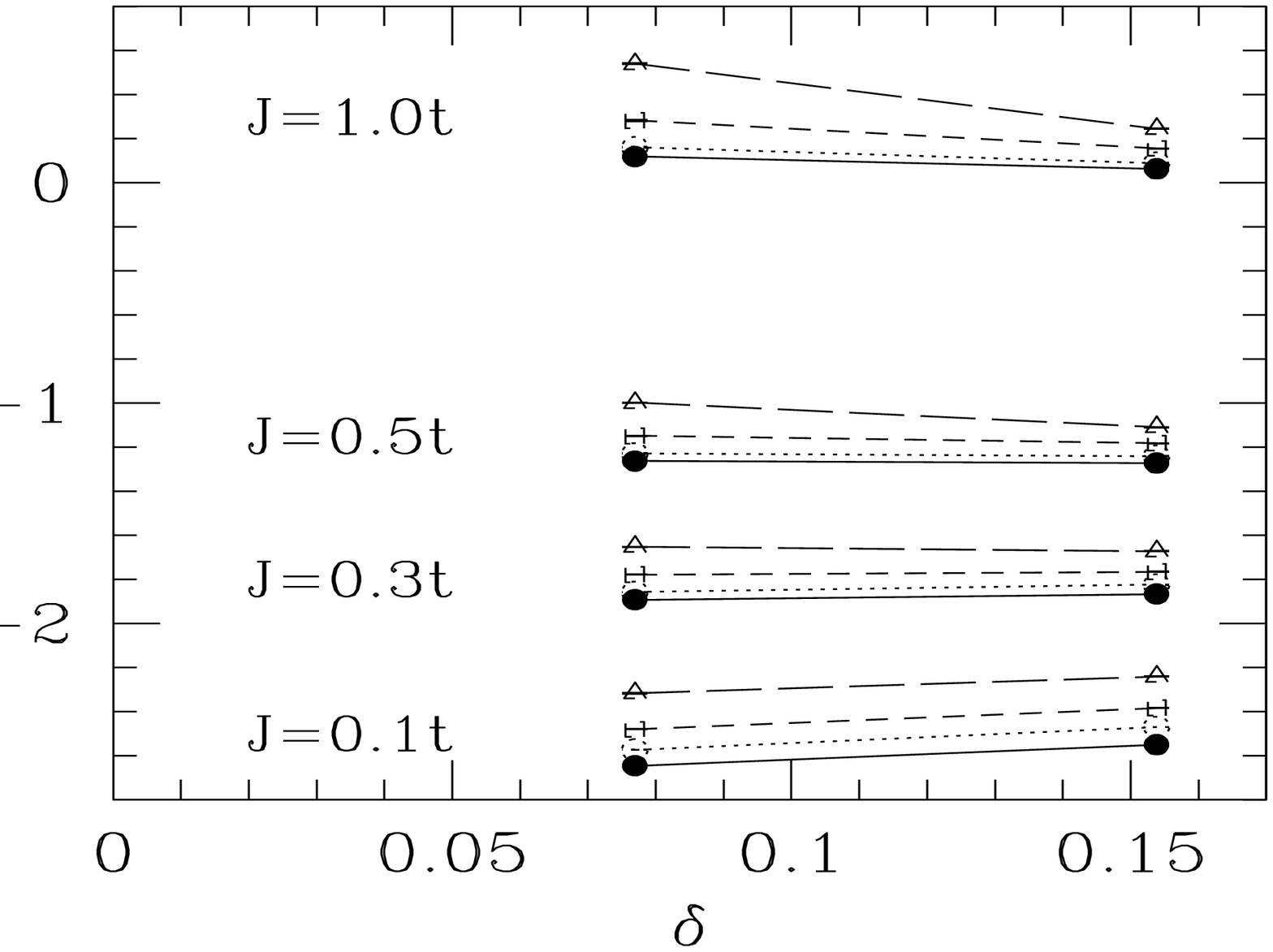,width=6.5cm}}
\caption {\baselineskip .185in \label{fig:exl26}
$e_h(\delta)$ computed with VMC  (long dashed lines),
 FN (short dashed lines), 
GFMCSR (dotted lines) compared with the exact results 
 \protect\cite{poilblanc} (full lines) for a $26$ lattice size.
  Error bars are much smaller
 than the size of the symbols.  Lines (guides to the eye) connect 
 the two and four hole results.}
\end{figure}
A fundamental ingredient in GFMC  is the choice of the guiding function in
order to perform importance sampling, as described in \cite{calandra}. 
At finite doping
we have used a pure d-wave BCS  guiding
wave function \cite{gros,giamarchi,note1} plus a long range density-density
Jastrow factor \cite{franjio}.
At  half filling instead we use  the 
guiding function described  in \cite{calandra} 
that allows to obtain the exact answer for $e_0$,
 as there is no sign problem at zero doping for this particular 
guiding function. In the following we use the exact $e_0$ at the given size 
$L$ for all variational (VMC), FN and GFMCSR calculations of $e_h(\delta)$.
Further we employ   periodic boundary conditions tilted by $45$  ($\sim 11$)
degrees on  square lattices with $L=2 l^{2}$   and odd integer 
$l\ge5$ ($L=26$).   

\begin{figure}
\centerline{\psfig{figure=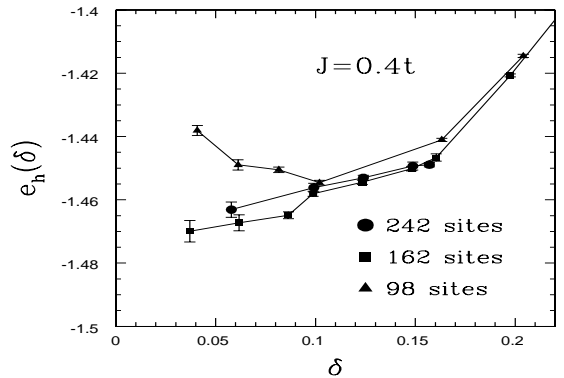,height=5.0cm}}
\caption {\baselineskip .185in\label{fig:phasefn}
$e_h(\delta)$ computed with FN for several lattice sizes.
}
\end{figure}

 We have compared our Monte Carlo calculations with the exact Lanczos 
results for the largest size $L=26$ available in the
 literature\cite{poilblanc}.
We have found that the FN 
approximation improves the ground state energy of the best 
starting variational (and guiding) 
 wavefunction by a factor of three (Fig.~\ref{fig:exl26}) 
and the GFMCSR by another similar factor, 
yielding finally an accuracy of less than 100K  on the  energy per hole, 
which is physically acceptable if compared  with  the low energy
 coupling of the model $J\sim 1500K$. 
\begin{figure}
\centerline{\psfig{figure=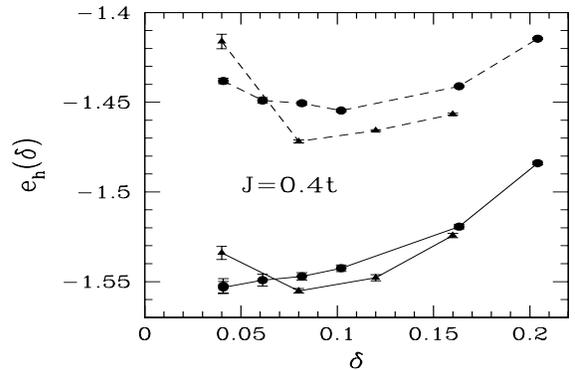,height=5.0cm}}
\caption {\baselineskip .185in \label{fig:phastoc}
$e_h(\delta)$  computed with FN (dashed lines) and with GFMCSR
(continuous lines) for $L=50$ (triangles) and $L=98$ (circles) at $J=0.4t$.
 Lines are guides to the eye.}
\end{figure}
As shown in  Fig. \ref{fig:exl26} this kind of accuracy depends weekly 
on the number of holes. However, for small lattices,    the
main difficulty to detect PS is the resolution in doping.
By increasing the system size (see Fig. \ref{fig:phastoc}), 
the difference between FN calculation and the GFMCSR one 
remains of the same order, and much below the VMC energies.
 Thus we expect that the accuracy 
of the calculation is not very much size dependent, even for large systems
where no exact solution is available. 
 We remark that,
 as  the accuracy of the calculation is improved  from FN to GFMCSR,
 the minimum in the hole energy disappears for the largest size in
 Fig.~\ref{fig:phastoc}.
This   suggests that 
the occurrence of PS at $J/t=0.4$ and at  this system 
size is an artifact of the FN. 
This  approximation naturally enhances
the tendency of PS because the FN acts only on 
the kinetic part of the Hamiltonian, thus implying a tendency to localize the 
holes.  Nonetheless, as shown in Fig. \ref{fig:phasefn},  we have obtained, 
even within the FN, the stability 
of the uniform phase in the thermodynamic limit at $J/t=0.4$.
Thus we conclude that PS at this $J/t$
 value is only {\em a finite size effect}.
Several authors \cite{lee,manousakis} have used the infinite $L$ limit for 
$e_0$ even for 
 the finite $L$ evaluation  of $e_h(\delta)$. We have instead   
 used  the exact $e_0$ for each lattice  size, and checked  that 
 for the largest size calculations both choices of $e_0$ lead 
 to the same $e_h(\delta)$. 
 Indeed  the small lattice size results 
 are very sensitive at low doping to the particular 
choice of $e_0$, and this  may explain the 
contradictory results presented  in the literature so far. 
Moreover  in Ref.\cite{manousakis} the most important  and delicate 
 low doping region is studied only with fairly small lattice sizes. 

The absence of PS is also confirmed by the 
low momenta behavior of $N(q)$, which will be discussed in the remaining 
part of this letter. 
In order to compute the 
ground state correlation functions $ N(q) $ 
  we have used two different methods: 
the  ''forward walking technique'' (FW) which allows the direct evaluation 
of the ground state expectation value, 
at the expense of  very large error bars 
when the convergence to the ground state is particularly  slow, 
as shown  in \cite{calandra}; the second technique is based on 
 Hellmann-Feynman theorem, and amounts to compute the ground state 
energy  $E(\lambda)$ of the Hamiltonian in presence of a small perturbation
  $\lambda N(q)$, $ N(q) = {d \over d \lambda } E(\lambda)\mid_{\lambda=0}$,
 the derivative 
being estimated numerically by a few runs for different $\lambda$'s. 
The latter  technique is much more stable, especially for large size, 
but  each $q$ value requires  several simulations,
 whereas  a single one is sufficient  for the 
FW technique for all $q$'s. Thus  we have used the more expensive method   
for the small $q$ values where the FW convergence is more difficult, and 
we have checked the consistency of both methods  in the remaining 
momentum region. Moreover the FW technique is limited to the FN 
approximation and has not been extended within  the GFMCSR scheme yet.
For the 26 site cluster the FN results for $N(q)$ are accurate within 
$3\% $, as compared with the exact diagonalization data.  
Since this accuracy is already very good for determining the qualitative 
features of $N(q)$, we have chosen to work within FN, avoiding to implement 
the much more expensive GFMCSR on large clusters. 

\begin{figure}
\centerline{\psfig{figure=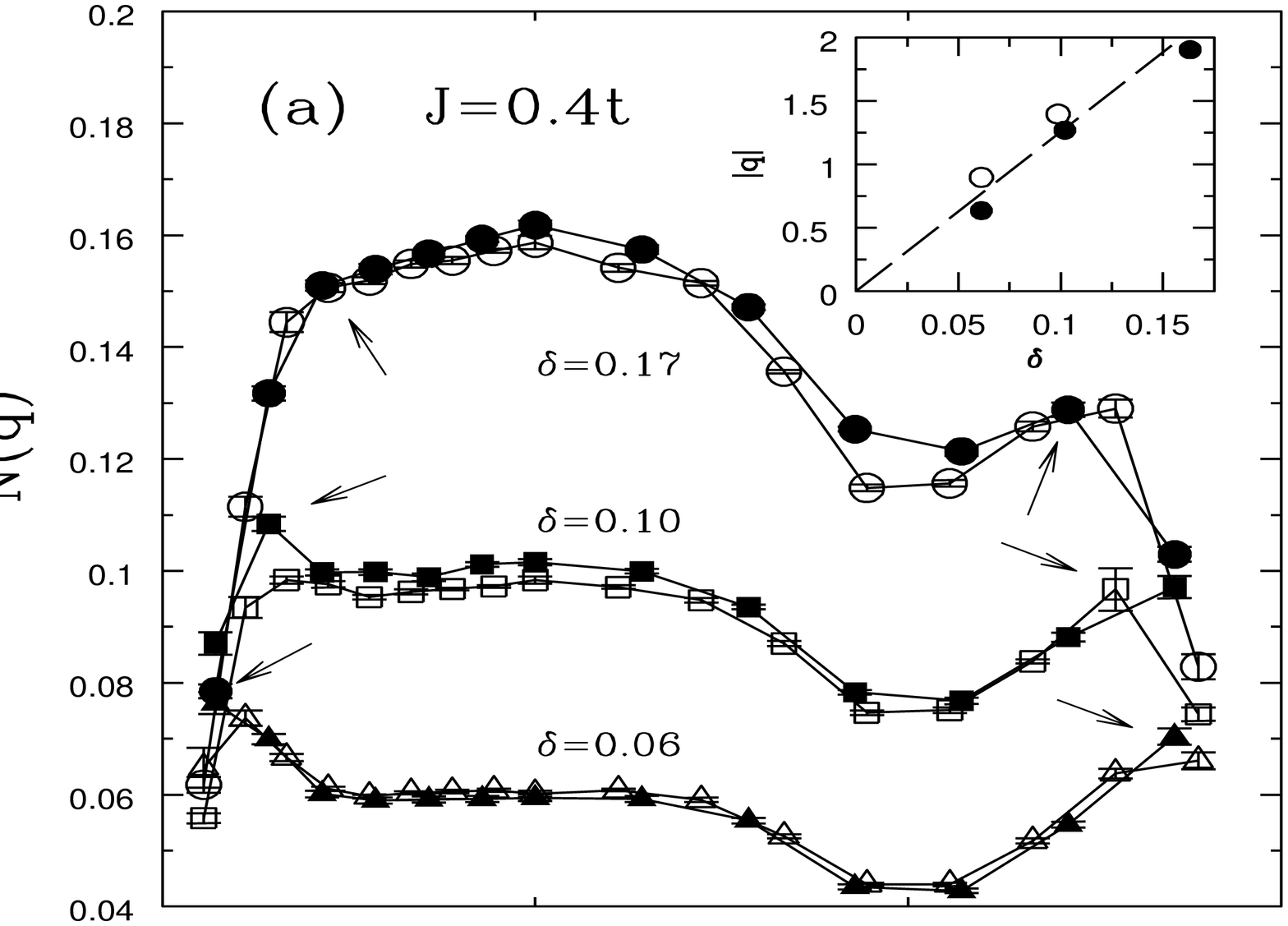,height=5.0cm}}
\centerline{\psfig{figure=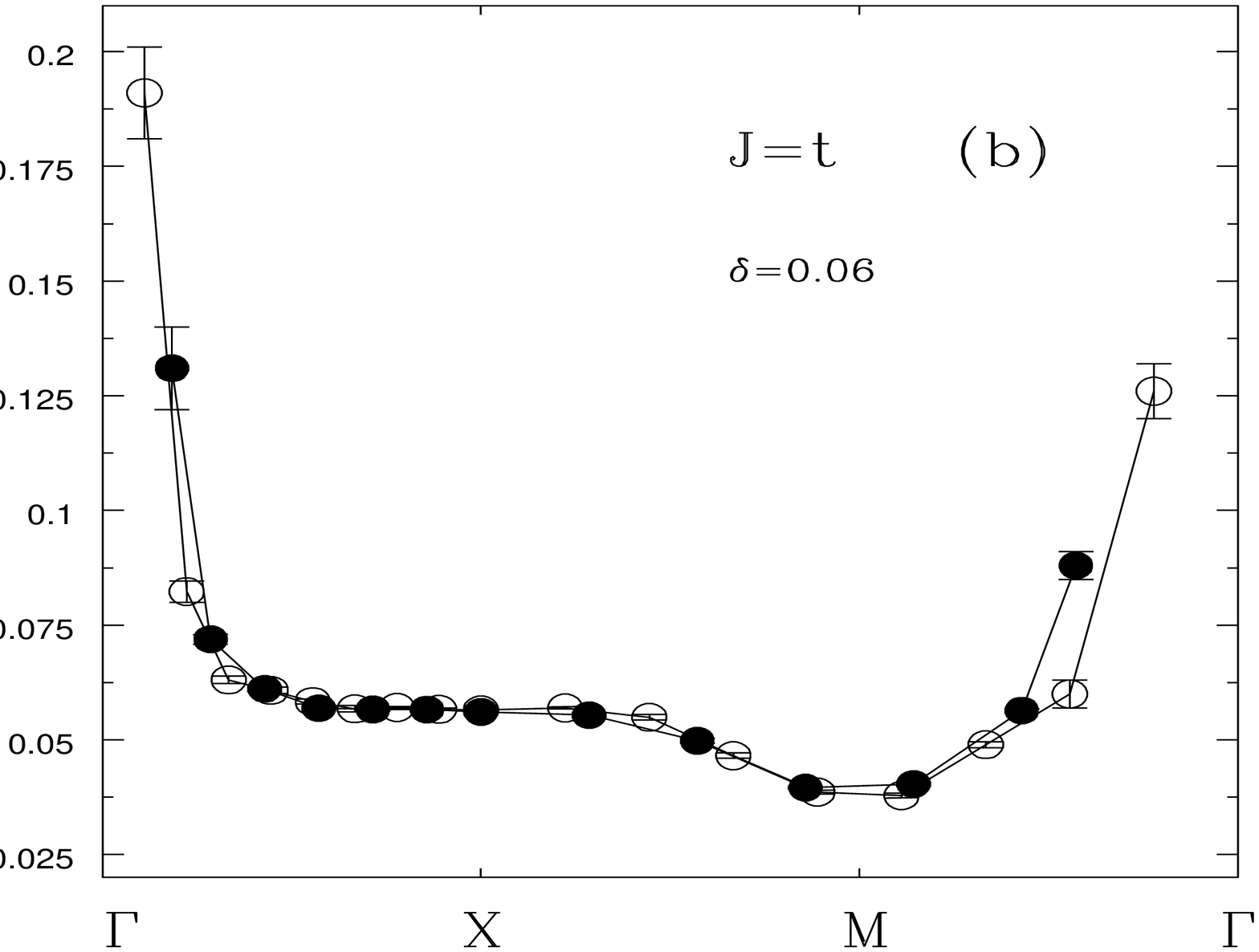,height=5.0cm}}
\caption {\baselineskip .185in\label{fig:nq}
(a) $N(q)$ for several doping $\delta=0.06$ (triangles), $\delta=0.10$
(squares), $\delta=0.17$ (circles) for different sizes, $L=98$ (full symbols)
and $L=162$ (empty symbols). Incommensurate
peaks are shown by the arrows. $\Gamma=(0,0)$, $X=(\pi,\pi)$, $M=(\pi,0)$.
 Inset: empty dots (full dots) represent
peak positions along  the $(1,0)$ direction in $q$ space (diagonal direction).
 The dashed line displays   the $4 \pi$ slope. (b) $N(q)$ for $\delta=0.06$
and $L=162$ (empty dots), $L=98$ (full dots)}
\end{figure}

In Fig.~\ref{fig:nq}  the $N(q)$ is plotted for several 
dopings and lattice sizes. 
 Inside the PS  region, at $J=t$ (Fig.~\ref{fig:nq} b),  
 $N(q)$ shows a  divergent peak at small $q$, 
as can be expected from general arguments.
  Instead for $J=0.4t$  (Fig.~\ref{fig:nq} a) 
the charge correlations  approach zero as $q \to 0$, 
confirming the absence of PS even at the lowest doping.
Nevertheless  enhanced fluctuations are clearly evident along 
the $(1,1)$ and $(1,0)$ directions. 
We believe that these incommensurate peaks   are  a genuine feature of the 
ground state of the model, because they do not appear for instance 
at the VMC level, and  it  is extremely important to use many 
 power iterations (\ref{iter})   
to   eliminate the  bias due to the VMC guiding wavefunction.  
 The $N(q)$ at this $J/t$ value is very weakly size dependent, so that 
the overall shape of this function in the thermodynamic limit should not 
differ too much from the   one shown in Fig.~\ref{fig:nq} (a).  
Thus $N(q)$ should be always finite even for small $q$, ruling out PS 
and charge density wave instability. Even though we found some 
peaks at incommensurate wavevectors, that maybe 
reminiscent of some dynamical stripe order, 
 they are not consistent with 
a static stripe structure  (the peaks should  diverge).
In fact we tried to stabilize some static stripe order similarly  to 
what was done  within Density Matrix Renormalization Group (DMRG)\cite{white},
 but in our approach the uniform phase remains always the most stable one
 at $J/t=0.4$.    The stabilization  of the stripe phase with DMRG is 
probably due to the use of open boundary condition in one direction.  
 We have not attempted to use open boundary condition as,
i) the   momentum is not defined on finite lattices,  ii)  
it is more difficult to reach the thermodynamic limit,  iii) the 
 gap to the first excited state is much reduced, especially close 
to a PS  instability, yielding a much slower convergence of the power  
method  (\ref{iter}). 

Remarkably, as $\delta$ is increased the peak at finite momentum moves far 
from the $\Gamma$ point  at a distance that scales linearly with the doping 
with a coefficient which is surprisingly close to $4 \pi$. 
This is exactly the coefficient obtained experimentally in 
$La_{2-x-\delta}Nd_{x}Sr_{\delta}CuO_{4}$ \cite{yamada}.
We  have also found that   
this  peak position does not depend on $J/t$, implying
that the $ 4 \pi \delta $ slope could be 
a general feature of the $t-J$ model in the region without PS. 
It is reasonable to expect that the interaction of electrons with the ions 
could further enhance the intensity of this incommensurate peaks, 
leading not only to a qualitative but also to a quantitative agreement 
with experiments\cite{tranquada,yamada}. 

In conclusion we have performed accurate calculations 
of the ground state energy  of the  2D $t-J$ model 
with  three  methods: VMC, FN and GFMCSR. 
\begin{figure}
\centerline{\psfig{figure=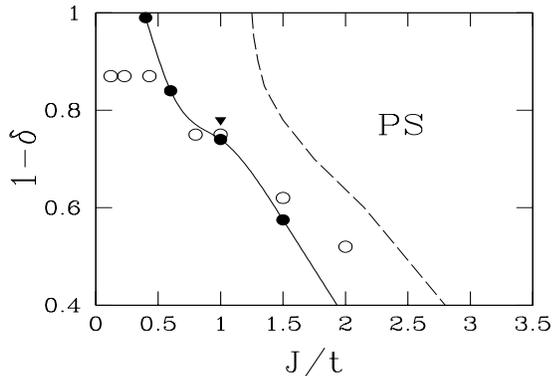,height=5.0cm}}
\caption {\baselineskip .185in\label{fig:diagram}
The low doping PS phase diagram of the $t-J$ model. 
Full dots (connected by continuous line) are computed with FN, the 
full triangle with GFMCSR, empty dots are from \protect\cite{emery}
and the dashed line is from \protect\cite{rice}.}
\end{figure}
As  summarized in the phase diagram picture (Fig.~\ref{fig:diagram}) 
we find  no evidence of PS for $J\le0.4 t$, at least for $\delta \ge 2.5\%$, 
which represents the smallest doping considered here ($6$ holes in a $L=242$ 
sites), and a transition to the phase separated regime at a critical $J_c \sim  0.5t$.
The small doping difference between FN and GFMCSR for the occurrence of
PS at $J=t$ strengthens the validity of the FN phase separation diagram.
These  results are in acceptable quantitative 
agreement with Ref. \cite{lee,kohno,heeb}
but we believe that our calculation represents a  much better  
attempt to control the finite size effects, which  are very important, 
especially at small doping.  
Instead, in the large doping region,  the best   agreement  is found   
with the exact diagonalization data\cite{emery}  on small clusters.
Close to PS we have computed the equal time charge correlations and enhanced
fluctuations at incommensurate momenta have been found. 
The fact that  the position of these peaks  approach the $\Gamma$ 
point linearly with the doping  cannot be explained within a  simple 
model  for the holes, like e.g. the spinless fermion one, as was  
 proposed  by \cite{putikka} to explain the shape of $N(q)$ using HTE. 
In this case in fact the characteristic incommensurate wavevector 
$2k_F$ approaches the $\Gamma$ point in a much more singular fashion 
$2 k_F \sim \sqrt{\delta}$. On the other hand the hard-core boson 
model is unable to produce any feature at momenta different from $\Gamma$.
Further study is probably necessary to clarify this point. 
We emphasize however that, within the $t-J$ model,  it is possible 
to reproduce the qualitative features  of the experimental findings of 
incommensurate charge fluctuations, 
without i) electron-phonon coupling 
 and  ii)   long 
range forces  to push the $q=0$ PS instability to finite $q$ 
value\cite{castellani}.

We  acknowledge D. Poilblanc for  the exact diagonalization 
$L=26$ data  and E. Heeb for 
 sending  us a copy of his PHD thesis.  We also thanks T. M. Rice, 
J. Gubernatis, L. Capriotti, M. Fabrizio, A. Parola, and W. Putikka 
  for useful discussions. 
This work was supported in part by INFM (PRA HTSC) and CINECA grant. 


\begin{references}
\bibitem{note1} We have checked that,  by adding an AF 
order parameter  to the wavefunction,  
as  considered in Ref. \cite{gros,giamarchi}, the VMC energy improves 
but not  the $FN$ and $GFMCSR$ ones.
\end{references}
\end{document}